\documentclass[sn-mathphys-num]{sn-jnl}
\usepackage{graphicx}%
\usepackage{multirow}%
\usepackage{amsmath,amssymb,amsfonts}%
\usepackage{amsthm}%
\usepackage{mathrsfs}%
\usepackage[title]{appendix}%
\usepackage{xcolor}%
\usepackage{textcomp}%
\usepackage{manyfoot}%
\usepackage{booktabs}%
\usepackage{algorithm}%
\usepackage{algorithmicx}%
\usepackage{algpseudocode}%
\usepackage{listings}%
\usepackage{cleveref}

\theoremstyle{thmstyleone}%
\theoremstyle{thmstyletwo}%

\theoremstyle{thmstylethree}%

\begin{document}

\title[Article Title]{A Bayesian approach to model uncertainty in single-cell genomic data}

\author*[1]{\fnm{Shanshan} \sur{Ren}}\email{shanshan.ren.21@ucl.ac.uk}

\author[1]{\fnm{Thomas E.} \sur{Bartlett}}\email{thomas.bartlett.10@ucl.ac.uk}

\author[2]{\fnm{Lina} \sur{Gerontogianni}}\email{linager94@gmail.com}

\author[3]{\fnm{Swati} \sur{Chandna}}\email{s.chandna@bbk.ac.uk}

\affil[1]{Department of Statistical Science, University College London, London, U.K.}
\affil[2]{Bioinformatics and Biostatistics Science Technology Platform, The Francis Crick Institute, London, U.K.}
\affil[3]{School of Computing and Mathematical Sciences, Birkbeck, University of London, London, U.K.}

\abstract{Network models provide a powerful framework for analysing single-cell count data, facilitating the characterisation of cellular identities, disease mechanisms, and developmental trajectories. However, uncertainty modeling in unsupervised learning with genomic data remains insufficiently explored. Conventional clustering methods assign a singular identity to each cell, potentially obscuring transitional states during differentiation or mutation. This study introduces a variational Bayesian framework for clustering and analysing single-cell genomic data, employing a Bayesian Gaussian mixture model with Dirichlet priors to estimate the probabilistic association of cells with distinct clusters. This approach captures cellular transitions, yielding biologically coherent insights into neurogenesis and breast cancer progression. The inferred clustering probabilities enable further analyses, including Differential Expression Analysis and pseudotime analysis. We propose the Area Under the Curve as a novel way of quantifying clustering performance in relation to cell-type specific marker genes, as it better captures the clustering characteristics of interest and aligns closely with visual assessments in the scRNA-seq data. To summarise, this methodological advancement enhances the resolution of single-cell data analysis, enabling a more nuanced characterisation of dynamic cellular identities in development and disease. }


\maketitle

\section{Introduction}\label{sec1}

The utilisation of network models in single-cell count data provides a novel and powerful approach for understanding complex cellular data. A network graph $G$ consists of a set of vertices $V$ and edges $E$, represented as $G = (V, E)$, which can be depicted using an adjacency matrix $A$. In an \textit{n}-node graph characterised by a singular node type (unipartite network), the adjacency matrix $A \in \{0, 1\}^{\mathit{n} \times \mathit{n}}$ is defined such that  $\forall i \in \{1, 2, ..., n\}$ and $j \in \{1, 2, ..., n\}$, $A_{ij} = 1$ if there exists an edge between node $i$ and $j$, and $A_{ij} = 0$ otherwise. If a set of nodes can be partitioned into two disjoint groups $U, V$ with no edges connecting vertices within $U$ or within $V$, then these nodes can be represented as a bipartite graph e.g as depicted in \cref{fig:adj} on page \pageref{fig:adj}---this represents a network graph $G$ comprising two distinct types of nodes. The adjacency matrix is \( A \in \{0, 1\}^{{p} \times n} \), where $p$ and $n$ are the number of nodes in set $U$ and set $V$ respectively. In the context of the count data framework, $p$ denotes the number of genes and $n$ is the number of cells, with the association between node $i \in \{1, 2, ..., p\}$ and node $j \in \{1, 2, ..., n\}$ represented by non-negative integer counts. 

Raw genomic sequencing data typically produces discrete counts of mRNA copies, which are generally represented as non-negative integers. The representation of such data as a matrix can be expressed as ${\mathbb{Z}}^{\ p\times n}_{\geq 0}$. Unique molecular identifiers (UMIs) are a widely utilised technique in single-cell sequencing data. They are expected to model raw counts through a multinomial distribution \cite{townes2019feature}, although this is not obligatory, offering a more precise representation of the data and alleviating duplication bias. Traditional quantification methods like TPM (Transcripts Per Million) and FPKM (Fragments Per Kilobase of Exon per Million Fragments Mapped) require normalisation, which can distort UMI data and introduce artificial correlation among transcripts \cite{townes2019feature}.

Due to limitations in experimental techniques, inferring cellular identities from genomic data has become a widely studied area in computational biology. Louvain clustering is a prevalent technique for grouping cells into discrete clusters in genomic data analysis, especially in single-cell RNA sequencing, as it optimises the modularity within a precomputed \textit{k}-nearest-neighbour (\textit{k}-NN) graph representation of the data. The Louvain clustering method is implemented in an R package named Seurat. The Seurat-Louvain clustering transforms the \textit{k}-NN graph into a more robust SNN graph by calculating the shared nearest neighbour (SNN) similarity among mutual neighbours of the cells derived from the \textit{k}-NN graph, resulting in a symmetric adjacency matrix, i.e., $A \in \{0, 1\}^{n \times n}$. As an alternative, we explicitly model the raw counts of mRNA molecules. We represent these counts as an asymmetric adjacency matrix $ A\in {\mathbb{Z}}^{\ p\times n}_{\geq 0}$, providing a biologically more accurate representation. 

An alternative method for understanding network block structures is through a probabilistic framework utilising the blockmodels. These models constitute a class of latent variable models \cite{hoff2002latent}\cite{young2007random}\cite{rubin2022statistical}. For each node in a network, it is associated with a latent categorical position, and the probability of an edge between two nodes is determined by their respective latent classes. In stochastic blockmodel (SBM), the probability of a link or edge between nodes i and j is defined as $P_{ij} = P_{ji} = B_{z_i z_j} \quad \forall i, j = 1,2, ...,n$, where $z_i$ denotes the block membership of vertex $i$. The matrix $B$ has dimensions $k \times k$, with $B_{ab} \in [0, 1] \quad \forall a, b$ \cite{holland1983stochastic}.  The limitation of the SBM is its assumption that all vertices possess an identical expected degree within the same block.  To address the degree heterogeneity of network graphs, the degree-corrected stochastic blockmodel (DC-SBM) is preferable, as it generalises the stochastic blockmodel (SBM) by introducing an additional parameter $\theta$ to account for the heterogeneous degrees of each node within the same block, thereby improving the accuracy in modelling real-world complex biological networks\cite{qin2013regularized}.  Furthermore, constraining the DC-SBM to execute a modularity-like approximation simplifies the optimisation problem of DC-SBM to a modularity maximisation.  Newman has demonstrated that maximising modularity is mathematically equivalent to a variant of spectral clustering utilising the modularity matrix \cite{newman2006modularity}. Under general network models, nodes are known to have latent positions with certain properties that are revealed in their graph spectral embeddings. Furthermore, these latent positions are known to be asymptotically multivariate Gaussian distributed in the spectral embeddings \cite{rubin2022statistical}. By integrating this fact with the alternating cellular identities in biological development, we propose a novel methodology that combines a Bayesian framework with Gaussian mixtures to model uncertainty during cell progression; specifically the Variational Bayesian estimation of a Gaussian Mixture Model (VB-GMM)\cite{blei2006variational}. In our model, distinct blocks in the blockmodel and components of the multivariate Gaussian mixture represent different cell types. This approach allows probabilistic cluster assignments, enabling the detection of transitional or ambiguous cellular states, which are common in processes such as cell differentiation or tumour progression. Our proposal is particularly valuable for identifying dynamic or rapidly replicating cells whose identities may be unclear using traditional clustering approaches. Finally, evaluating clustering performance in the absence of labels remains a significant challenge. We introduce the misclustering rate as a novel, quantitative metric for assessing the quality of cluster assignments in scRNA-seq data. This metric will be applied to benchmark datasets to validate the effectiveness of our proposed clustering methodology.

This paper is organised as follows:  In the \hyperref[sec:results]{Results} section, we present the clustering results obtained from our innovative method, the VB-GMM, juxtaposed with the Gaussian Mixture Model (GMM), both visually and quantitatively, applied to the embryo cortical dataset and the breast cancer dataset. In the \hyperref[sec:methods]{Methods} section, we detail the VB-GMM methodology and introduce the Area Under the Curve (AUC) as performance metrics. Finally, a \hyperref[sec:discussion]{Discussion} section is provided.\\

\section{Results}\label{sec:results}

\subsection*{The breast cancer dataset }
\begin{figure}
    \centering
    \includegraphics[width = 85mm]{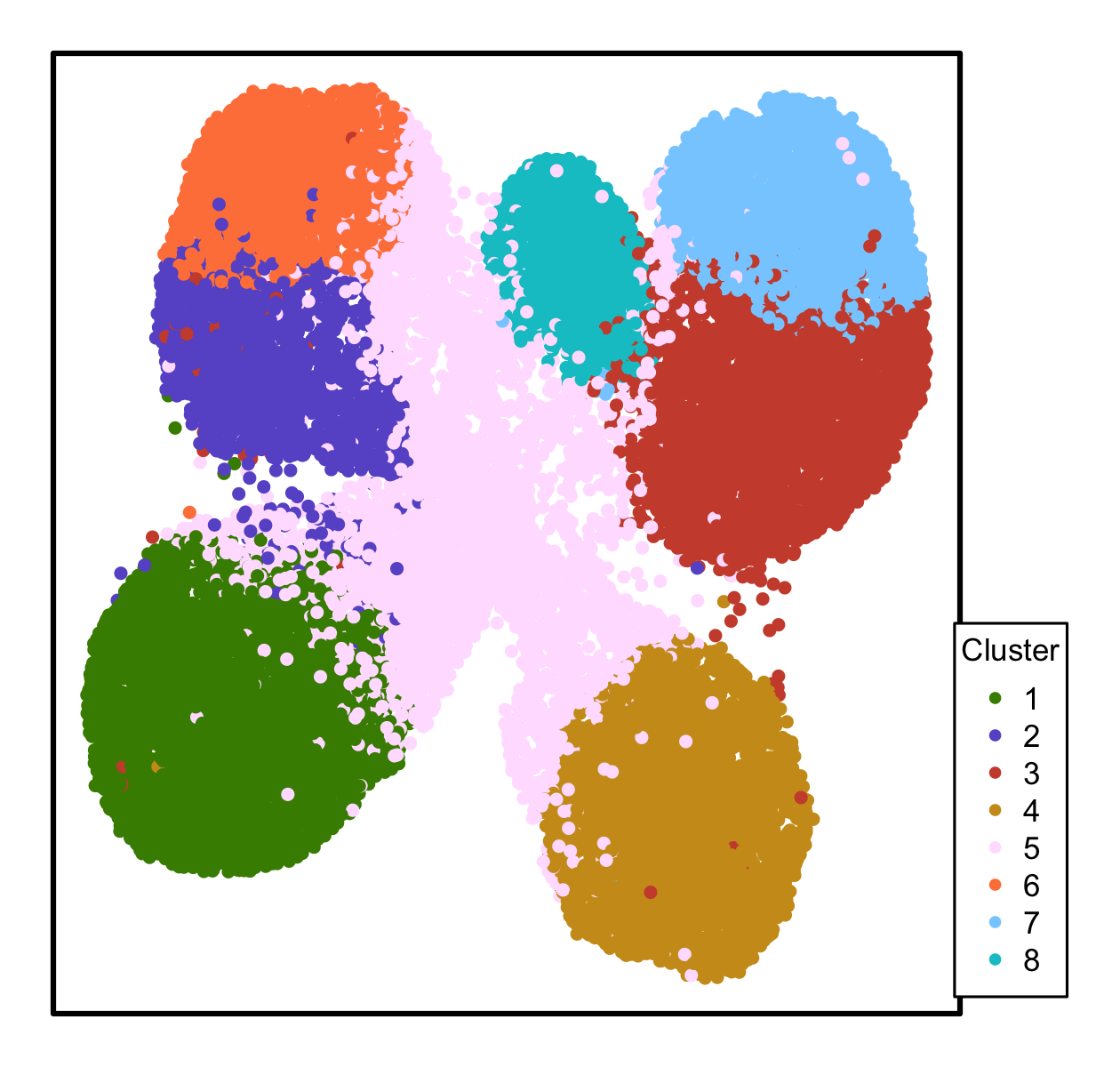}
    \caption{\textbf{GMM clusters for breast cancer data in UMAP-LE projection.}}
    \label{fig:GMM k8 BR}
\end{figure}

\begin{figure}
    \centering
    \includegraphics[width = 0.97\textwidth]{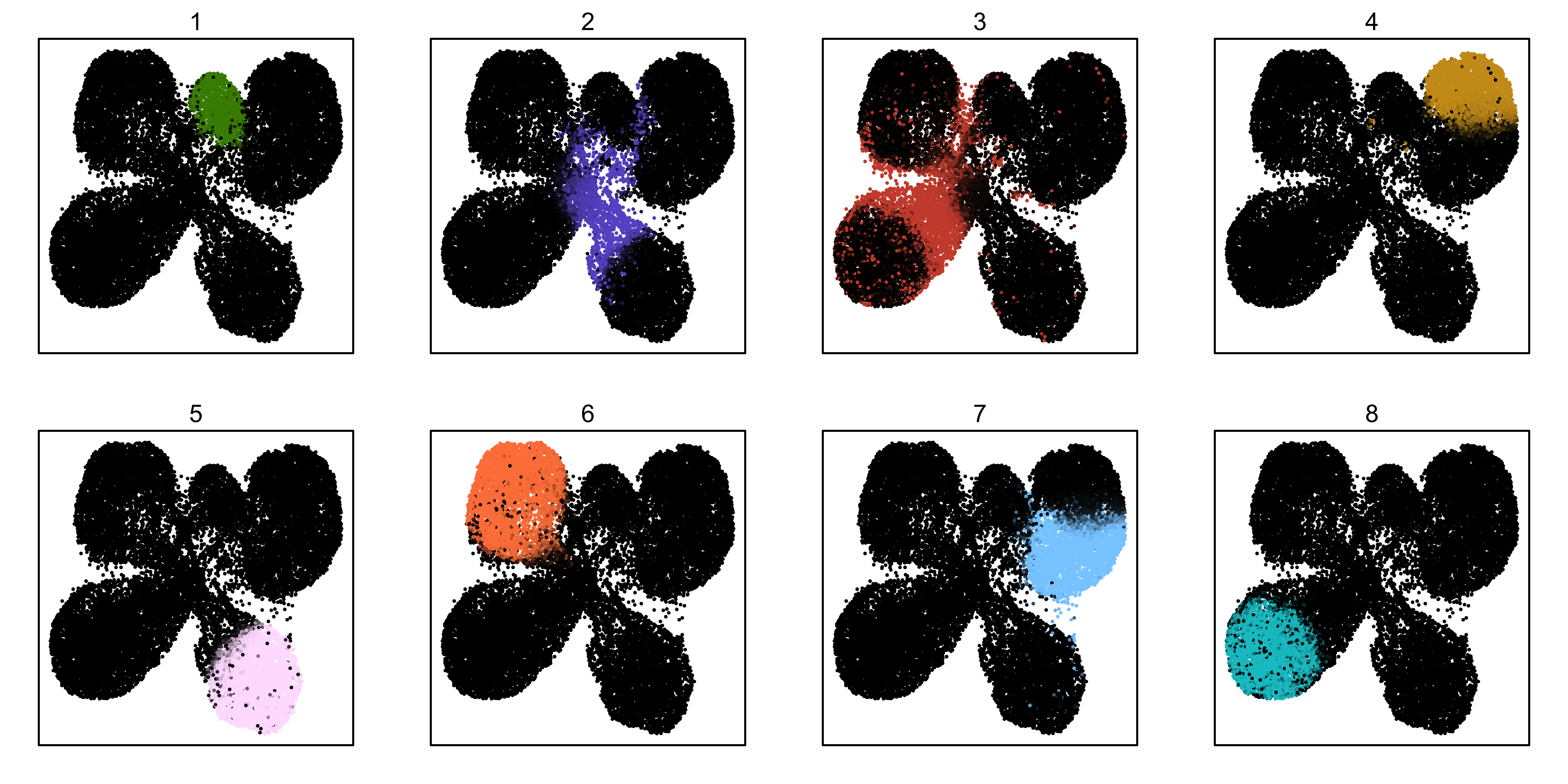}
    \caption{\textbf{VB-GMM clusters for breast cancer data in UMAP-LE projection.}}
    \label{fig:VI k8 BR}
\end{figure}

\begin{figure}
    \centering
    \includegraphics[width = 0.97\textwidth]{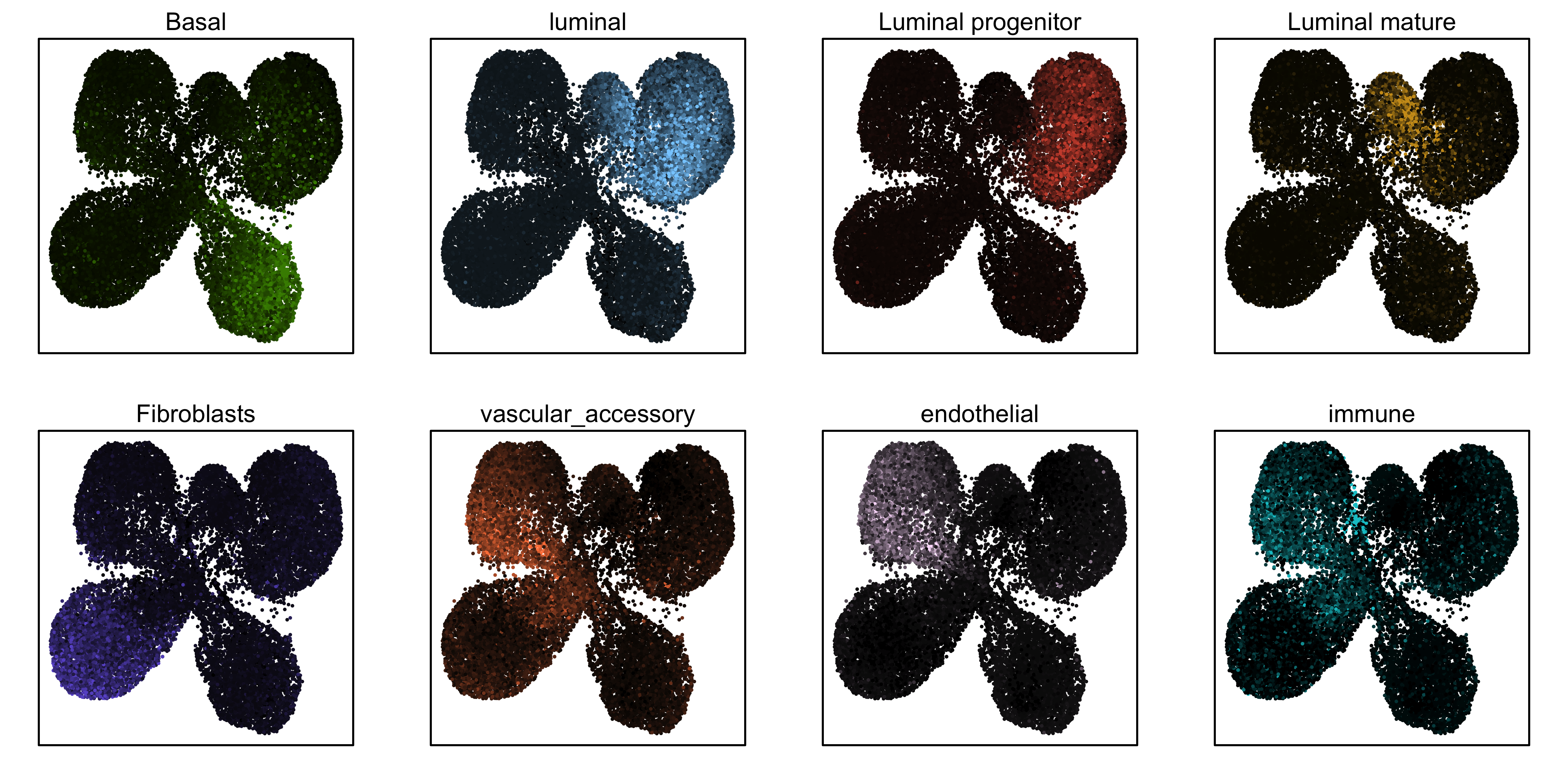}
    \caption{\textbf{Marker gene plot for breast cancer data in UMAP-LE projection.}This is a validation plot that shows the mean expression levels of targeted cell types.}

    \label{fig:LE marker BR}
\end{figure}

In breast cancer research, the cellular origin of tumours is critical. Hormone receptor-positive (HR+) breast cancers are typically derived from luminal mature cells, whereas hormone receptor-negative (HR-) cancers, including Triple-Negative Breast Cancer (TNBC), often originate from luminal progenitor cells\cite{tharmapalan2019mammary}. Therefore, comprehending these cell types is essential for investigating the causes of breast malignancies. The number of clusters $K_{br}$ evaluated with the breast cancer dataset is $K_{br} \in \{2, 3, 4, 5, 6, 7, 8, 10, 12, 14, 22, 25\}$. The computation commenced at $K=8$ for both GMM and VB-GMM, corresponding to the eight predefined cell types proposed by the biologists. For $K=8$, both GMM (Figure \ref{fig:GMM k8 BR}) and VB-GMM (Figure \ref{fig:VI k8 BR}) demonstrate a distinct separation of each cluster when shown in the marker plot (Figure \ref{fig:LE marker BR})---particularly in effectively clustering luminal progenitor and luminal mature cells. $K=8$ represents the optimal visualisation plots for both GMM and VB-GMM. When $K_{br}<8$, we forfeit some nuanced differentiation among various cell types; conversely, when $K_{br}>8$, we increasingly encounter clusters that contain noise or superfluous data (see \cref{fig:VI k6 BR}, \ref{fig:VI k22 BR}, \ref{fig:GMM k6 BR}, \ref{fig:GMM k22 BR} ).

It is noteworthy that both GMM and VB-GMM (Figure \ref{fig:GMM k8 BR}, \ref{fig:VI k8 BR}) further subdivided the region representing luminal progenitor cells in the marker plot (Figure \ref{fig:LE marker BR}) into two distinct clusters, cluster $3$ and cluster $7$ for GMM; cluster $4$ and cluster $7$ for VB-GMM. The segregation of luminal progenitor cell clusters may provide valuable insights and warrants additional investigation. We started the investigation by evaluating these clusters via applying the weighted SigClust algorithm \cite{keefe2025powerful}, which confirms that the two clusters of luminal progenitor are statistically different (p-value $<0.01$). The hypotheses made by the weighted SigClust algorithm here are $H_0$: all observations from cluster $7$ and cluster $3$ in GMM come from a single multivariate Gaussian distribution; $H_1$: all observations from cluster $7$ and cluster $3$ in GMM do not come from a single multivariate Gaussian distribution. The same hypotheses are tested for cluster $4$ and cluster $7$ of the VB-GMM.

After confirming the distinctness of these segregated clusters, we performed Differential Expression Analysis (DEA) via edgeR package for these clusters, designating cluster $7$ as the control group and cluster $3$ as the treatment group (modified group) for GMM; cluster $4$ as the control group and cluster $7$ as the treatment group (modified group) for VB-GMM. The multiple testing correction in the DEA was performed using the False Discovery Rate (FDR) method, and results are reported as adjusted p-values.

We hypothesise that the healthy luminal progenitor cells will effectively differentiate into luminal mature cells, while the remaining cells at the bottom of the luminal progenitor cluster (cluster $3$ in GMM; cluster $7$ in VB-GMM) are likely at risk. The three highest-ranked genes for the DEA of VB-GMM are {\it KLF6, EMP1, RTN4} (see \cref{tab:Top25DEA_br} for the complete DEA results), all of which demonstrate strong statistically significant differences in expression levels between the compared groups (adjusted p-value $< 0.01$). The occurrence of positive Log2FC values shows upregulation of the specified genes in cluster $7$ relative to cluster $4$. These genes are all significant in relation to the principles of breast cancer\cite{ahmat2019pivotal}\cite{hatami2013klf6}\cite{pathak2018rtn4}. Nevertheless, the DEA of GMM has found another strong statistical expressed set of genes (adjusted p-value $< 0.01$), which are also relevant to the development of cancers. The three highest-ranked genes we obtained here are {\it TNFAIP6, TM4SF1, HLA-B} (see \cref{tab:Top25DEA_br_GMM}). Among these genes, {\it TNFAIP6} was upregulated in cluster $3$ relative to cluster $7$ with a positive Log2FC, whereas {\it TM4SF1} and {\it HLA-B} were downregulated with a negative Log2FC. These patterns are also consistent with the biological findings of the present study regarding cancer development, although detailed information on some of these genes remains limited. \cite{yang2024role}\cite{chen2022transmembrane}\cite{dhatchinamoorthy2021cancer}. Fluctuations in the expression levels of these genes may influence the behaviour of luminal progenitor cells, potentially contributing to early-stage breast cancer characterisation and diagnosis.\\

\subsection*{The embryo neurone dataset}
\subsubsection*{Comparison between the Louvain method, Gaussian mixture model (GMM) and variational Bayesian estimation of a Gaussian mixture (VB-GMM)} 

\begin{figure}[H]
    \centering
    \includegraphics[width = 105mm]{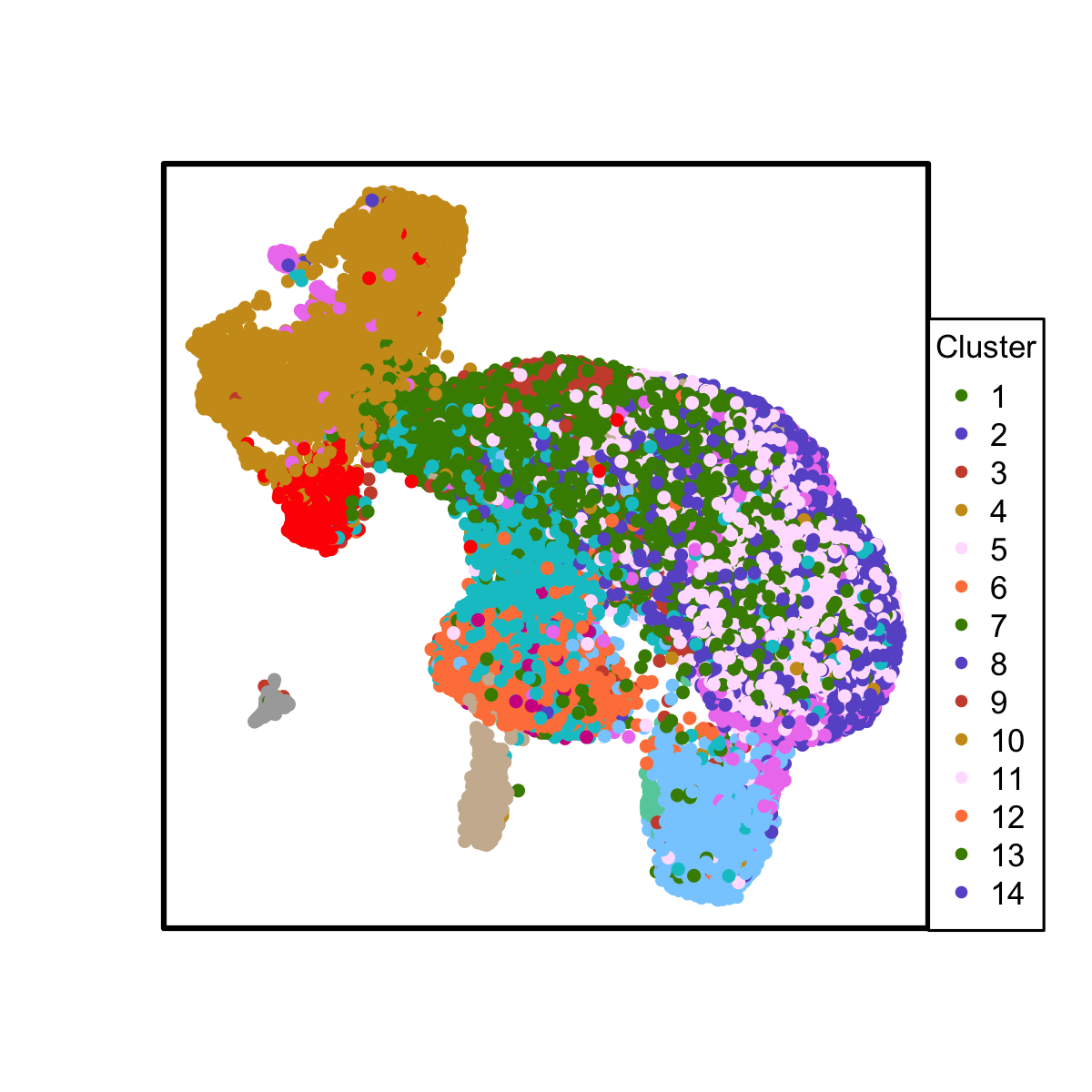}
    \caption{\textbf{The clusters of the Louvain Method for embryo data in UMAP-LE projection}}
    \label{fig:srt}
\end{figure}

\begin{figure}[H]
    \centering
    \includegraphics[width = 85mm]{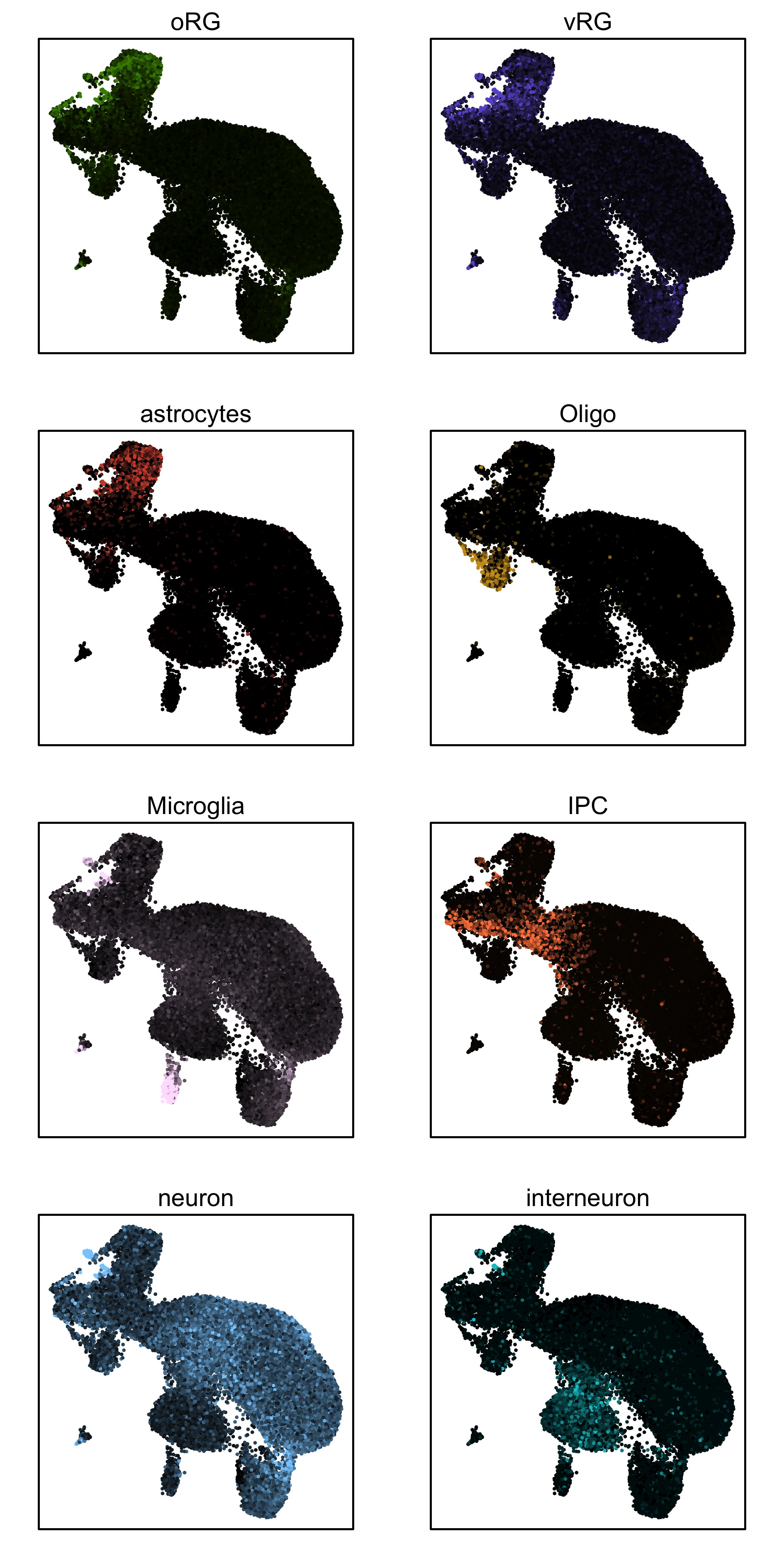}
    \caption{\textbf{Marker gene plot for embryo data in UMAP-LE projection.} This is a validation plot that shows the mean expression levels of targeted cell types.}
    \label{fig:LE marker gw1722 dbl}
    
\end{figure}

\begin{figure}[H]
    \centering
    \includegraphics[width = 105mm]{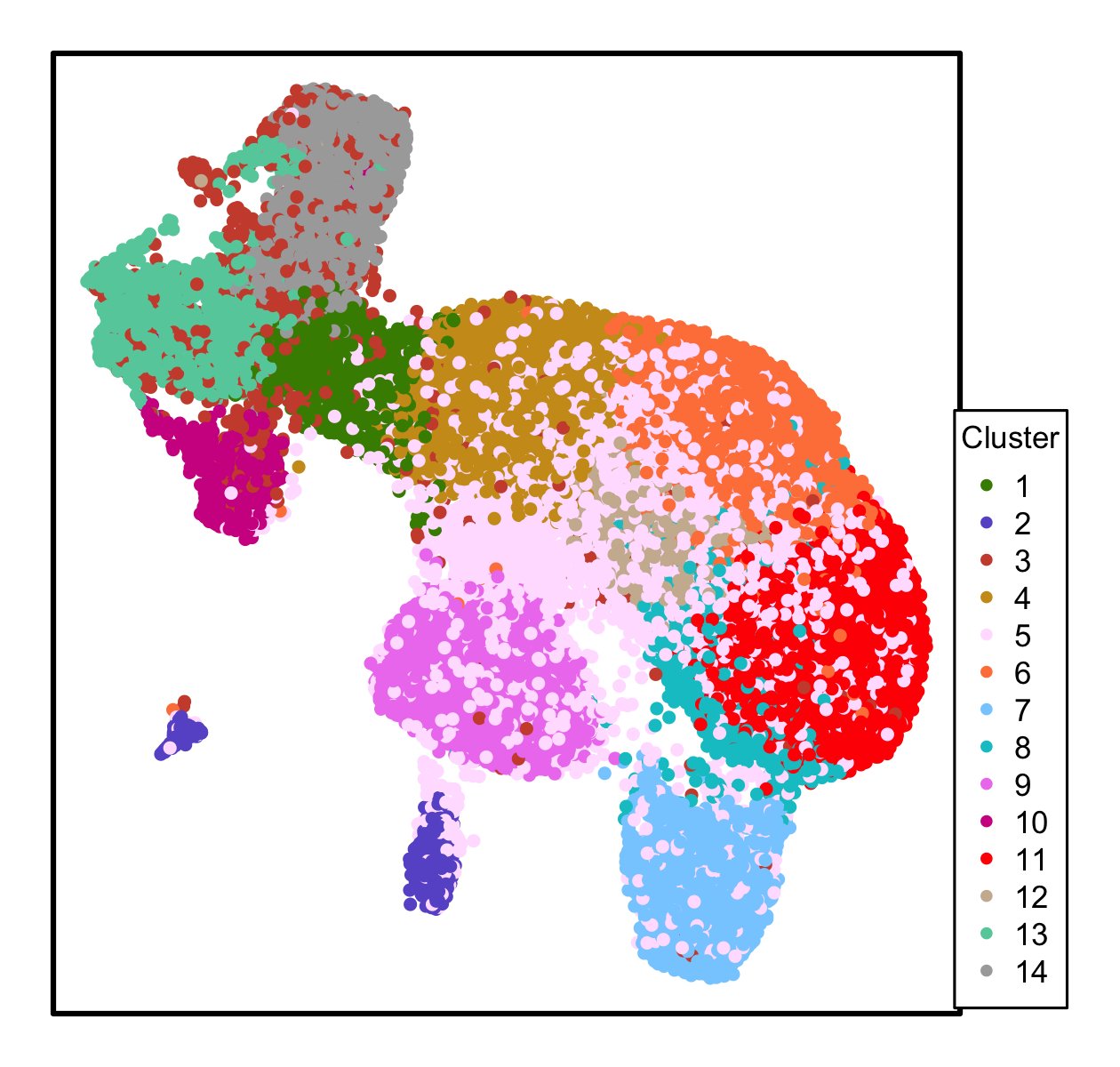}
    \caption{\textbf{GMM clusters for embryo data with $K=14$ in UMAP-LE projection}}
    \label{fig:GMM gw1722}
\end{figure}

\begin{figure}[H]
    \centering
    \includegraphics[width = 0.97\textwidth]{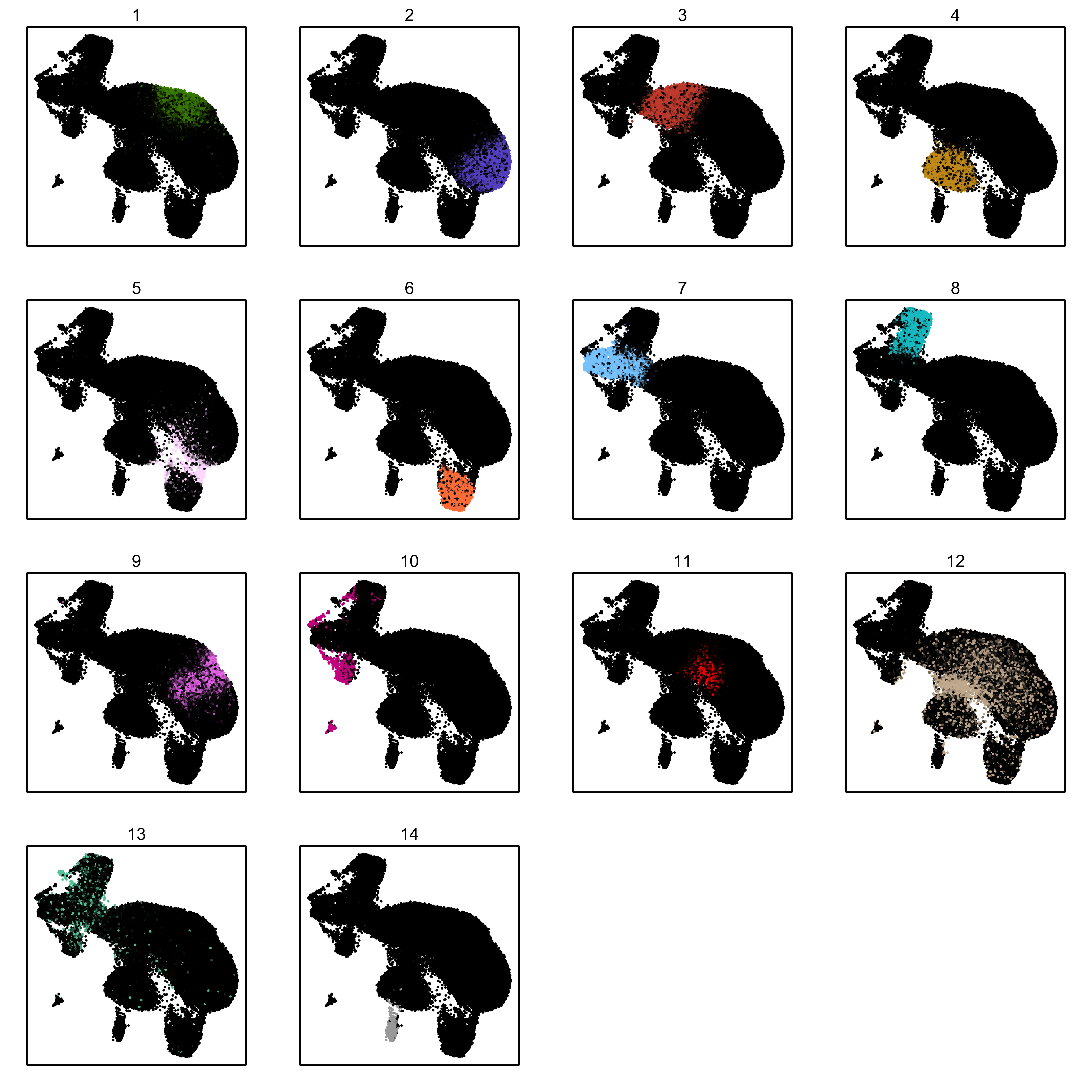}
    \caption{\textbf{VB-GMM clusters for embryo data with $K=14$ in UMAP-LE projection}}
    \label{fig:VB gw1722}
\end{figure}

The state-of-the-art method for cell clustering is the Louvain algorithm within the Seurat package. When comparing the Louvain plot (Figure \ref{fig:srt}) on page \pageref{fig:srt} with the marker plots (Figure \ref{fig:LE marker gw1722 dbl}) on page \pageref{fig:LE marker gw1722 dbl}, it demonstrates that the Louvain method effectively differentiated the fundamental structures of Oligo, Microglia, neuron, and interneuron. However, it fails to differentiate the cell types concentrated towards the upper part of the projection. Within this area, outer radial glia (oRG) and ventricular radial glia (vRG) are crucial cell types for the earliest phase of cortical cell formation in
the ectoderm, and hence it is of interest to see a clear separation between them. To keep the comparisons of the clustering efficiency among Louvain, GMM and VB-GMM consistent, $K=14$ is manually chosen for GMM and VB-GMM. When comparing the Louvain plot with the GMM plot (Figure \ref{fig:GMM gw1722}) on page \pageref{fig:GMM gw1722}, the GMM plot offers a more intricate representation of data segmentation. A clearer division is seen, especially with respect to the differentiation between oRG and vRG. GMM categories cluster $14$ as oRG, cluster $13$ as vRG, cluster $10$ as Oligo, cluster $2$ as Microglia, clusters $1$ and $4$ as intermediate progenitor cells (IPC), clusters $3, 6, 7, 8, 11$ and $12$ as neuron, and cluster $5$ and $9$ as interneuron. Compared with the GMM plot (Figure \ref{fig:GMM gw1722}), the VB-GMM plot (Figure \ref{fig:VB gw1722}) provides a clearer visualisation of individual cluster separation, including a more distinct separation between oRG and vRG cells. The posterior cluster assignment probabilities estimated by VB-GMM offer an alternative perspective on clustering results, enabling the quantification and consideration of assignment uncertainty. This provides additional information that could be used in downstream analyses, such as pseudotime analysis.

We noticed that identifying the astrocytes in the marker plot with either the GMM or VB-GMM plots is challenging. Andrews et al.\cite{andrews2022mechanisms} indicated that the oRG and vRG that did not differentiate into neurons instead developed into astrocytes. It is also important to note that the marker data originates from a distinct experiment compared to the embryo cortex data. Therefore, it is justifiable to omit astrocytes while analysing the clustering results.\\

\subsubsection*{Quantitative results}
To quantitatively evaluate clustering performance, we used AUC to assess clustering results from a different perspective. By treating cluster assignments as predictions of cell types and evaluating them using marker genes predefined by domain experts, this approach provides an alternative measure of clustering performance, particularly for scRNA-seq datasets lacking ground-truth labels.

\begin{figure}[H]
    \centering
    \includegraphics[width = 0.97\textwidth]{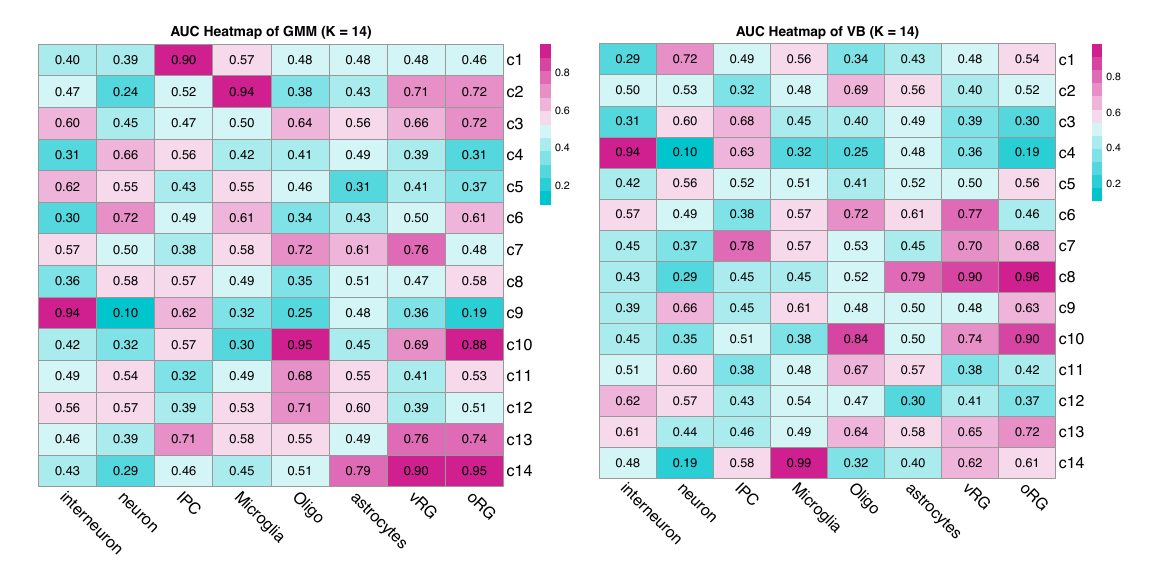}
    \caption{\textbf{AUC heatmap of GMM (left) and VB-GMM (right) on neuron data.} AUC values range from $0$ to $1$. Higher AUC values indicate superior discriminatory capacity, reflecting the probability that the classifier assigns higher scores to cluster $c_i$, $i \in \{1, 2, ..., 14\}$ than the rest of the clusters for each of the eight markers.}
    \label{fig:AUC}
\end{figure}

As shown in Figure \ref{fig:AUC}, the highest AUC score for each of the eight cell types is at least 0.70, with most exceeding 0.90. Several clusters exhibit high AUC scores for multiple cell types, including oRG, vRG, IPCs, astrocytes, neurons, and interneurons. Among these, it is particularly interesting that the clusters with the highest AUC score for astrocytes, cluster 14 for GMM and cluster 8 for VB-GMM, also exhibit high AUC scores for vRG and oRG. This observation is consistent with the biological evidence that astrocytes can arise from both vRG and oRG populations \cite{andrews2022mechanisms}. Similarly, the overlap among vRG, oRG, IPCs, and neurons may reflect their developmental relationships, as vRG can give rise to oRG and IPCs, which subsequently contribute to neuronal production.

In contrast, microglia exhibit their highest AUC scores in clusters that do not show similarly high AUC scores for vRG and oRG. This observation is also consistent with their distinct developmental origin, as microglia originate from yolk sac-derived primitive myeloid progenitors \cite{ginhoux2010fate}, whereas vRG and oRG are derived from the neuroepithelial progenitor cells during cortical development \cite{andrews2022mechanisms}.

We additionally performed Gene Set Enrichment Analysis (GSEA) and Differential Expression Analysis (DEA) on various clusters to ascertain whether the distinct clusters of the finer visualisation plot for VB-GMM ($K=14$) contain their respective gene sets and whether there are statistically significant differences in their expression levels. To answer that, multiple evaluations have been undertaken (see \cref{tab:GSEA_DEA_SIG}, \cref{tab:GSEA_DEA_SIG1}). An intriguing outcome arises from the examination of cluster $5$ (control group) and cluster $6$ (treatment group), which are distinct clusters, each representing neurones. Through the execution of Gene Set Enrichment Analysis (GSEA) and Differential Expression Analysis (DEA) utilising a gene set comprising marker genes for cell types found in human tissue, we uncovered notable genes such as {\it FABP7} and {\it DOK5}, which are implicated in neuronal and cerebral development \cite{de2012radial} \cite{pan2013dok5}. The DEA results for these genes indicated that cluster $6$ exhibits downregulation (negative log2FC) and demonstrates statistically significant differences in expression compared to cluster $5$ (adjusted \textit{p}-value $<0.01$). Subsequent evaluation of these clusters via weighted SigClust also shows significant statistical differences (\textit{p}-value $<0.01$) between these two clusters. Therefore, we possess greater confidence in our visualisation results, which support bigger values of $k$, than in the quantitative results, suggesting the plausible existence of finer clusters.\\


\section{Methods}\label{sec:methods}

\subsection*{Data pre-processing}
To begin with the data pre-processing, the raw count data was downloaded. We first filter out the empty cells, in which we keep a gene if it is expressed in at least 3 cells, and keep a cell if it has at least 200 genes expressed (non-zero values). Then, for the gene-level filtering, we exclude mitochondrial, ribosomal, and some irrelevant or dominating genes (``5S\_rRNA'', ``7SK",...etc.), and we retain genes expressed in at least 50 cells. For cell-level filtering, we retain cells with less than $10\%$ of their reads mapped to mitochondrial genes, cells where less than $50\%$ of reads are ribosomal, cells with at least 750 expressed genes, and cells where the most abundant gene makes up less than $10\%$ of total reads. Following cell-level filtration, we finally eliminated doublets using the scDblFinder package in R.
After downstreaming, we have:\\
\begin{enumerate}
    \item \textit{Breast cancer at-risk}\cite{pal2021single}: $n = 24284$ cells, $p = 16812$ RNA transcripts. This test dataset is from human breast tissue (healthy individuals).\\

    \item  \textit{Embryo cortical development}\cite{bhaduri2021atlas}: $n = 41734$ cells, $p = 16858$ RNA transcripts. This test dataset is from the V1 area of the visual cortex in human embryos' brains at gestation weeks 17-22.\\
    
\end{enumerate}

\subsection*{Specification of our statistical model}
By considering $A = X \in \mathbb{Z}_{\geq 0}^{p \times n}$\cite{bartlett2023stochastic}, the count matrix, where $i \in\{1,...,p\}$ represents genes in the data, $j \in\{1,...,n\}$ represents cells in the data, we can further model the observed distributions of genomic sequencing data in the eigenspace of the graph Laplacian by applying the regularised graph Laplacian $\mathbf{L_{\tau}}=\left(\mathbf{D_{\tau}}^{(X)}\right)^{-1/2}\mathbf{A}\left(\mathbf{D_{\tau}}^{(Y)}\right)^{-1/2}$, where $\mathbf{D_{\tau}}^{(X)} = \mathbf{D}^{(X)} + \tau_X I$, $\mathbf{D_{\tau}}^{(Y)} = \mathbf{D}^{(Y)} + \tau_Y I$, $\mathbf{D}^{(X)} = \sum_{j} A_{ij}$, and $\mathbf{D}^{(Y)} = \sum_{i} A_{ij}$. Then, by implementing the singular value decomposition (SVD), we obtain $L_{svd} = U \Sigma V^T$, where $U \in \mathbb{R}^{p \times d}$ and $V^T \in \mathbb{R}^{d \times n}$ orthogonal. We will subsequently employ $V$ for clustering to group cells exhibiting similar patterns of expression across genes.  The optimal dimensions of $V$ are selected according to the leverage score\cite{qin2013regularized}, defined as $\rVert V_j \rVert_{2}$ for each $j\in\{1, 2, ..., n\}$, where $V_j$ represents the $j^{th}$ row of $V$. $V$ will then be normalised before fitting the Variational Bayesian Gaussian mixture model for clustering. A Gaussian mixture model would be appropriate to conduct spectral clustering of the nodes, as the clusters of points in each latent position are asymptotically multivariate Gaussian\cite{rubin2022statistical}. \\
\subsection*{Variational Bayesian Estimation of a Gaussian Mixture \cite{blei2006variational}}
The Dirichlet distribution is used to model the mixture weights:
\begin{equation}
    \pi = (\pi_1, \pi_2, \dots, \pi_K) \sim \text{Dirichlet}(\alpha_1, \alpha_2, \dots, \alpha_K).
\end{equation}
For $v_{k} \subseteq V$, the likelihood is a mixture of Gaussians:
\begin{equation}
    p(v_k) = \sum_{k = 1}^K \frac{\alpha_k}{\sqrt{(2\pi)^d |\Sigma_k|}} exp[-\frac{1}{2}(v_k-\mu_k)^T \Sigma_{k}^{-1}(v_k-\mu_k)]
\end{equation}
Posterior samples of the mean vectors, the covariance matrix, and the weights for each component are $\boldsymbol{\mu}_k\in\mathbb{R}^{d}$, $\boldsymbol{\Sigma}_k\in\mathbb{R}^{d\times d}$, and $\pi_k\in\mathbb{R}$ respectively. Moreover, Eq.\ref{post} is the posterior estimate of the cluster assignment probability:
\begin{equation}
\hat{p}(k|j)=\frac{f_{\mathcal{N}}(\mathbf{v}^{(j)}|\boldsymbol{\mu}_k,\boldsymbol{\Sigma}_k)\cdot\pi_k}{\sum_{k'=1}^{K}f_{\mathcal{N}}(\mathbf{v}^{(j)}|\boldsymbol{\mu}_{k'},\boldsymbol{\Sigma}_{k'})\cdot\pi_{k'}},
\label{post}
\end{equation}
where $\mathbf{v}^{(j)}\in\mathbb{R}^d$.

\subsection*{Area Under the Curve (AUC)}
Given $t$ different cell types in marker gene scores, $k$ different clusters in the result of the clustering algorithm and $n$ number of cells, we have a $t$ by $n$ marker gene score matrix $M$ and the clustering labels $CL = \{CL_1, CL_2, ..., CL_n\}$. Then, the AUC between $M$ and $CL$ is
\begin{equation}
    AUC = \frac{(D_{M_i, CL_{k'}}+1)}{2}
\end{equation}
where $D_{M_i, CL_{k'}}$ is the Somers’ $D$\cite{newson2002parameters} rank correlation between $M_i$ and $CL_{k'}$, with $i \in \{1, 2, ..., t\}$ and $k' \in \{1, 2, ..., K\}$\\

\subsection*{Implementation details}
All analyses were performed using Python and R. To ensure reproducibility, the analyses have been re-executed using multiple versions of the required packages. The code used for data processing, model implementation, and result reproduction is publicly available at \url{https://github.com/ssren-sponge/VB-GMM}.

\section{Discussion}\label{sec:discussion}

In this paper, we presented a variational Bayesian clustering approach for scRNA-seq data that effectively models uncertainty in cell progression, offering deeper insights into dynamic cellular identities during development. We tested and validated that both VB-GMM and GMM methods perform well for clustering unsupervised scRNA-seq data, while the Bayesian method allows us to interpret the probabilistic assignment of cells to cell-types, thereby enabling us to model the uncertainty of data and gain dynamic information about the cells over time. 

This method has been applied to neurodevelopment and breast cancer datasets, yielding promising results by providing detailed cluster information, showing potential for further investigation. The efficacy of clustering algorithms applied to scRNA-seq data is typically assessed through the comparison of visual representations. To complement our analysis, we employed AUC as a quantitative metric, which provides a biologically informed perspective on clustering performance by assessing the association between clusters and cell types using predefined marker genes. This approach enables clustering evaluation without requiring definitive cell-type annotations, making it particularly valuable for scRNA-seq datasets in which cell-type boundaries are difficult to establish.
 
Another problem related to this research question is the automatic selection of the optimal number of clusters $K$ for clustering tasks---a longstanding challenge in the field of clustering that has garnered attention, and becomes increasingly difficult when dealing with extensive and high-dimensional complex count data. We aim to minimise this problem by methodically selecting $K$ through systematic testing, guided by visual inspection and correlation between cell clusters and known marker genes. 

In conclusion, our scRNA-seq clustering approach strengthened the connection between statistical theory and single-cell biology, offering a novel framework for modelling cellular dynamics and highlighting limitations of existing metrics in accurately evaluating clustering performance in complex biological networks.

\section*{Declarations}

\begin{itemize}
\item \textbf{Acknowledgements:}
The authors acknowledge the use of the UCL High Performance Computing Facilities, and associated support services, in the completion of this work.

\item \textbf{Ethics approval and consent to participate:}
Not applicable

\item \textbf{Consent for publication:}
Not applicable

\item \textbf{Availability of data and materials:}
The dataset of breast cancer supporting the conclusions of this article is available in the Gene Expression Omnibus under accession number GSE161529, \url{https://www.ncbi.nlm.nih.gov/geo/query/acc.cgi?acc=GSE161529}.\\
The dataset of embryo cortical development supporting the conclusions of this article is available from URL, 
\url{https://data.nemoarchive.org/biccn/grant/u01_devhu/kriegstein/transcriptome/scell/10x_v2/human/processed/counts/}.

\item \textbf{Code availability:}
The codes for our proposed method as implemented in this work are available online at: \url{https://github.com/ssren-sponge/VB-GMM} 

\item \textbf{Software:}
Downstream analyses were performed in R (versions 4.3–4.5), with the exception of VB-GMM, which was implemented in Python (version 3.11).

\item \textbf{Competing interests:}
The authors declare that there are no competing interests.

\item \textbf{Funding:}
This work was supported in part by the Research Innovation Fund awarded by Birkbeck, University of London, UK.

\item \textbf{Author contributions:}
All authors designed the study. SR wrote the manuscript. Statistical analyses
and model development were carried out by TB and SR. All co-authors
approved this version of this manuscript.

\end{itemize}

\bibliography{ref}

\break
\begin{appendices}

\section{Supplementary figures and tables}\label{secA1}

\begin{figure}[H]
    \centering
    \includegraphics[width = 85mm]{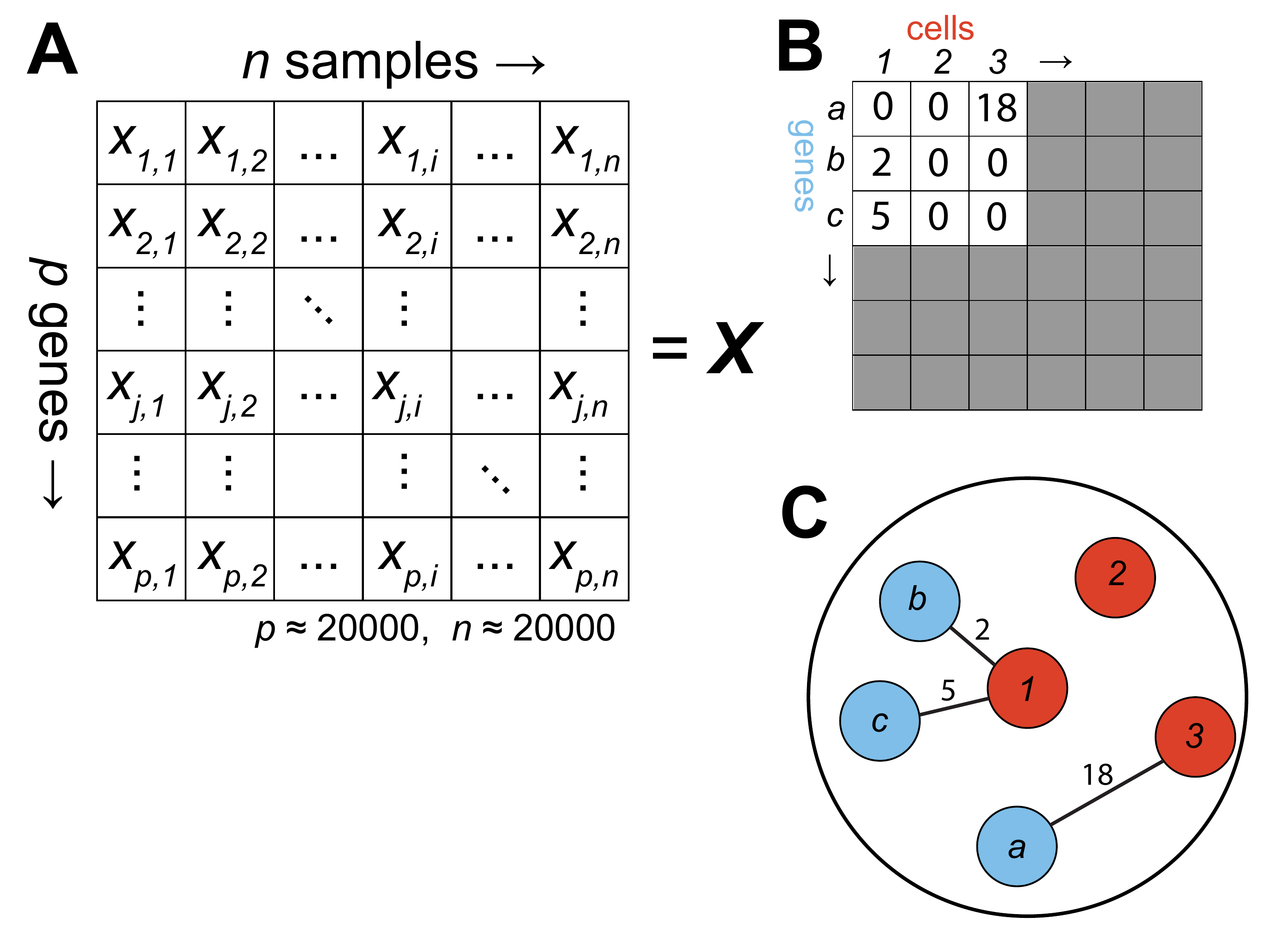}
    \caption{\textbf{Adjacency matrix $A  \in \{0, 1\}^{p \times n}$ of an asymmetric bipartite network, reproduced from Bartlett et al\cite{bartlett2023stochastic}.} \textbf{A} shows an asymmetric bipartite multi-edge network with adjacency matrix $A\in {\mathbb{Z}}^{\ p\times n}_{\geq 0}$; \textbf{B} models the data matrix $A$, comprising non-negative integer counts; \textbf{C} is the corresponding network of \textbf{B}. 
    }
    \label{fig:adj}
\end{figure}

\begin{figure}[H]
    \centering
    \includegraphics[width = 85mm]{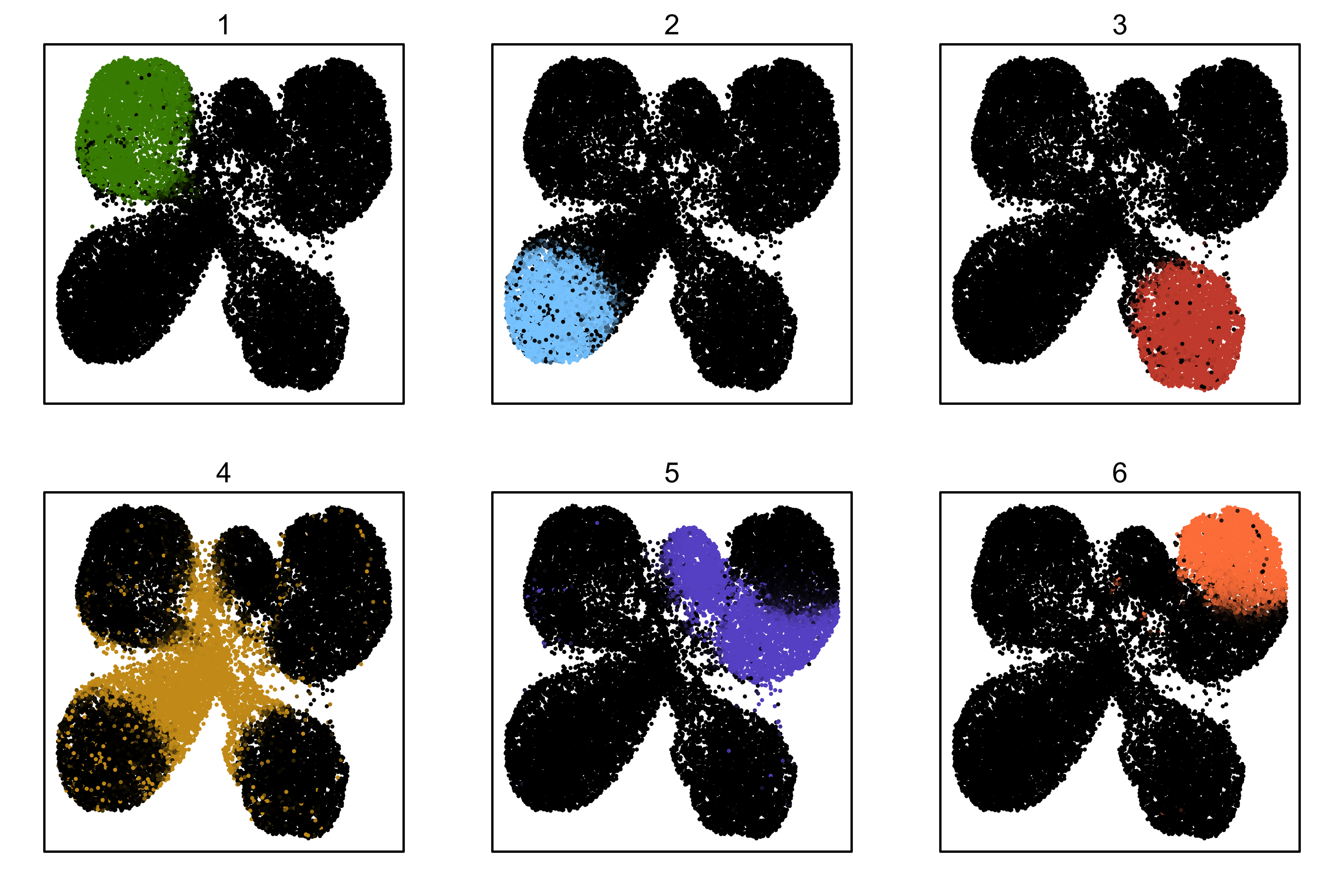}
    \caption{\textbf{VB-GMM clusters for breast cancer data with $k=6$ in UMAP-LE projection.}}
    \label{fig:VI k6 BR}
\end{figure}

\begin{figure}[H]
    \centering
    \includegraphics[width = 0.97\textwidth]{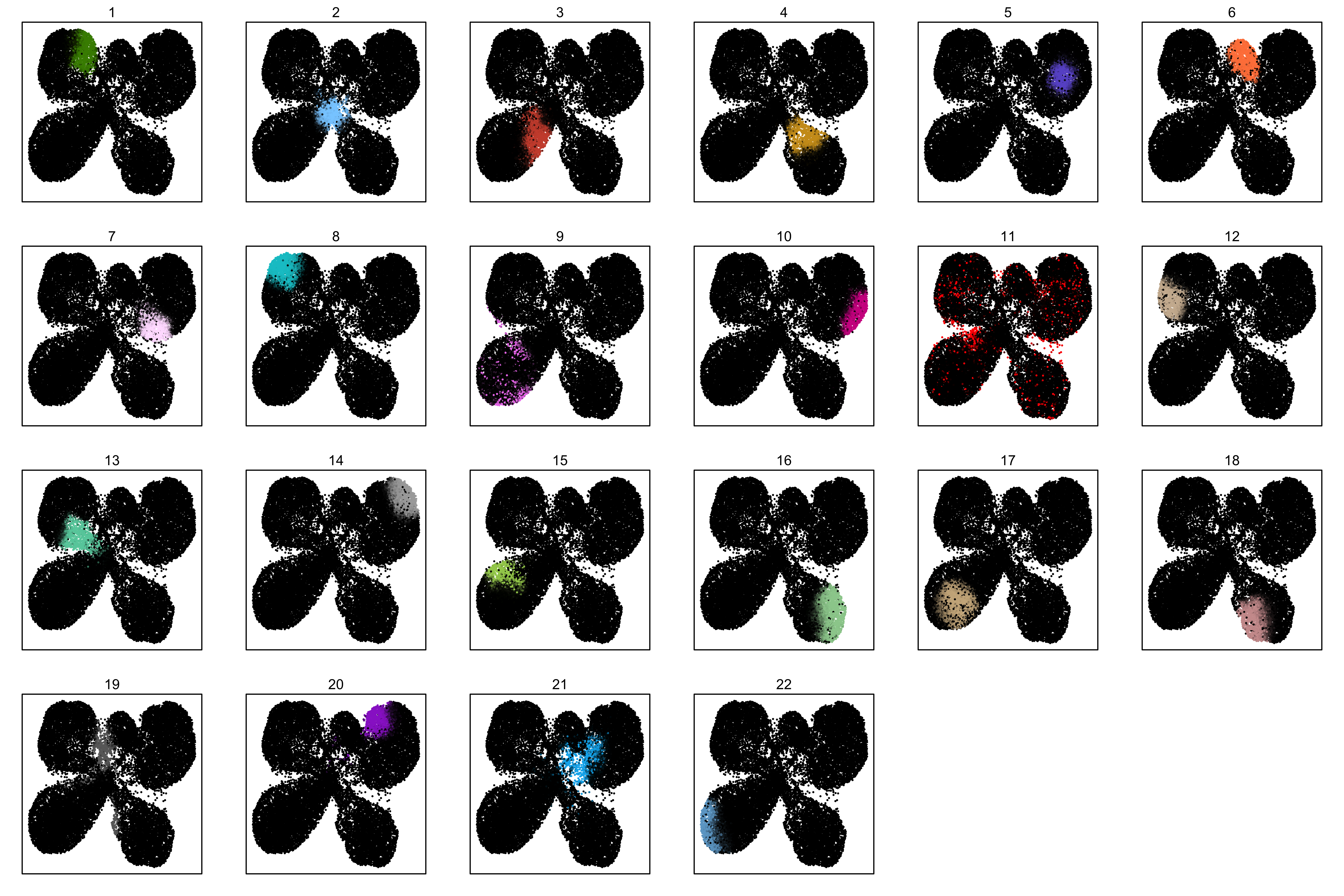}
    \caption{\textbf{VB-GMM for breast cancer data with $k=22$ in UMAP-LE projection.}}
    \label{fig:VI k22 BR}
\end{figure}

\begin{figure}[H]
    \centering
    \includegraphics[width = 85mm]{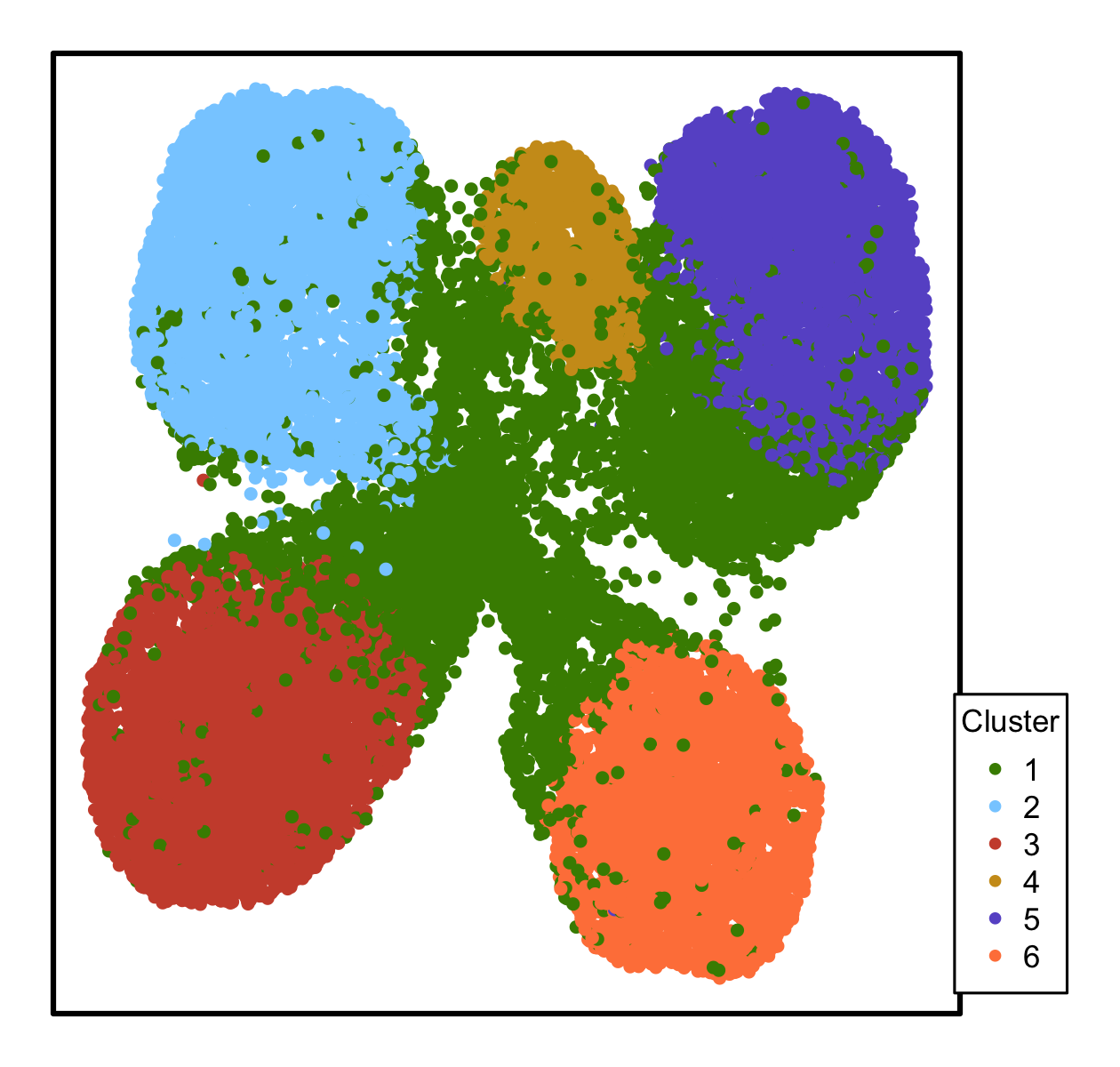}
    \caption{\textbf{GMM clusters for breast cancer data with $K=6$ in UMAP-LE projection.}}
    \label{fig:GMM k6 BR}
\end{figure}

\begin{figure}[H]
    \centering
    \includegraphics[width = 85mm]{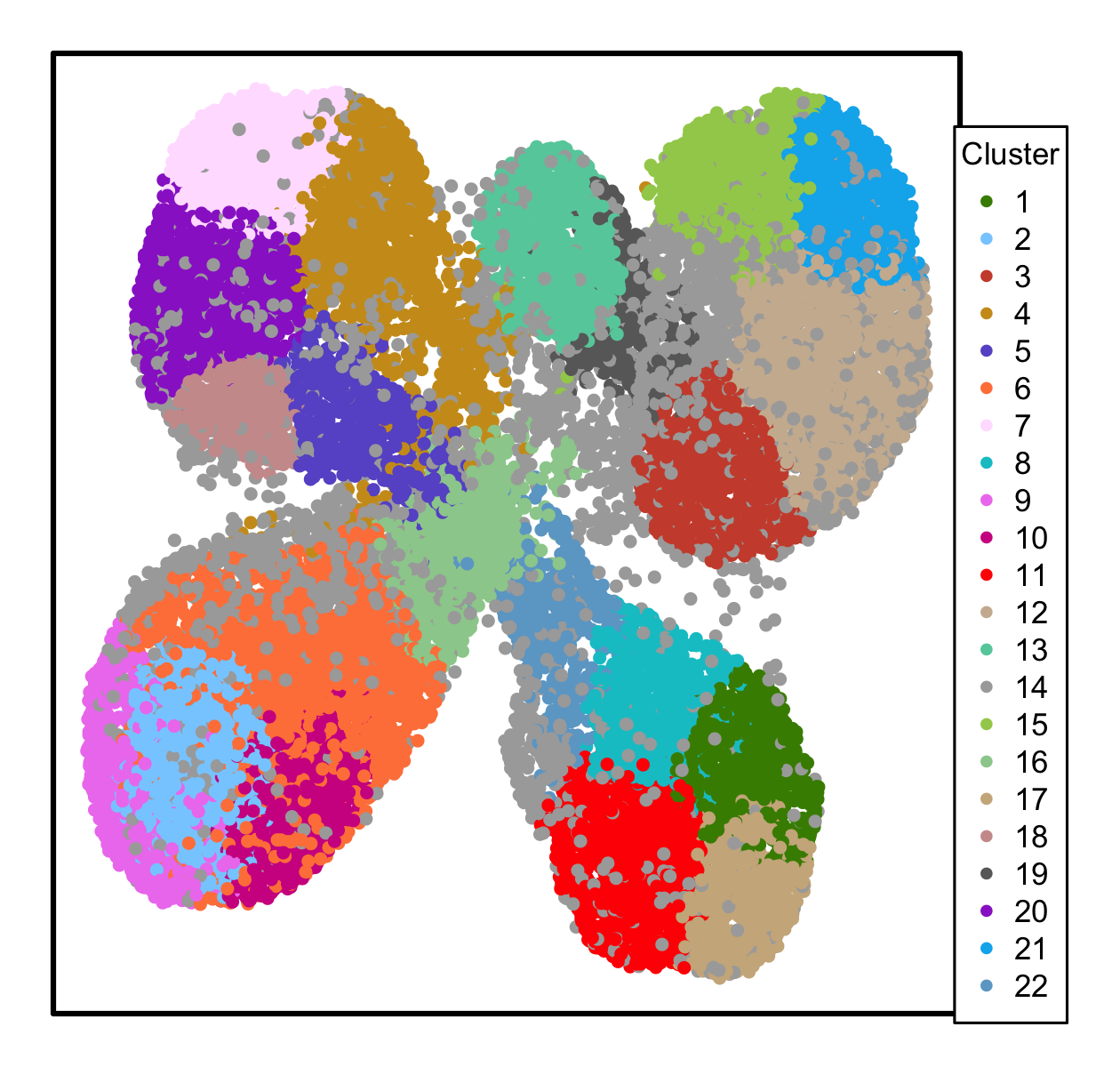}
    \caption{\textbf{GMM clusters for breast cancer data with $k=22$ in UMAP-LE projection.}}
    \label{fig:GMM k22 BR}
\end{figure}

\begin{figure}[H]
    \centering
    \includegraphics[width = 145mm]{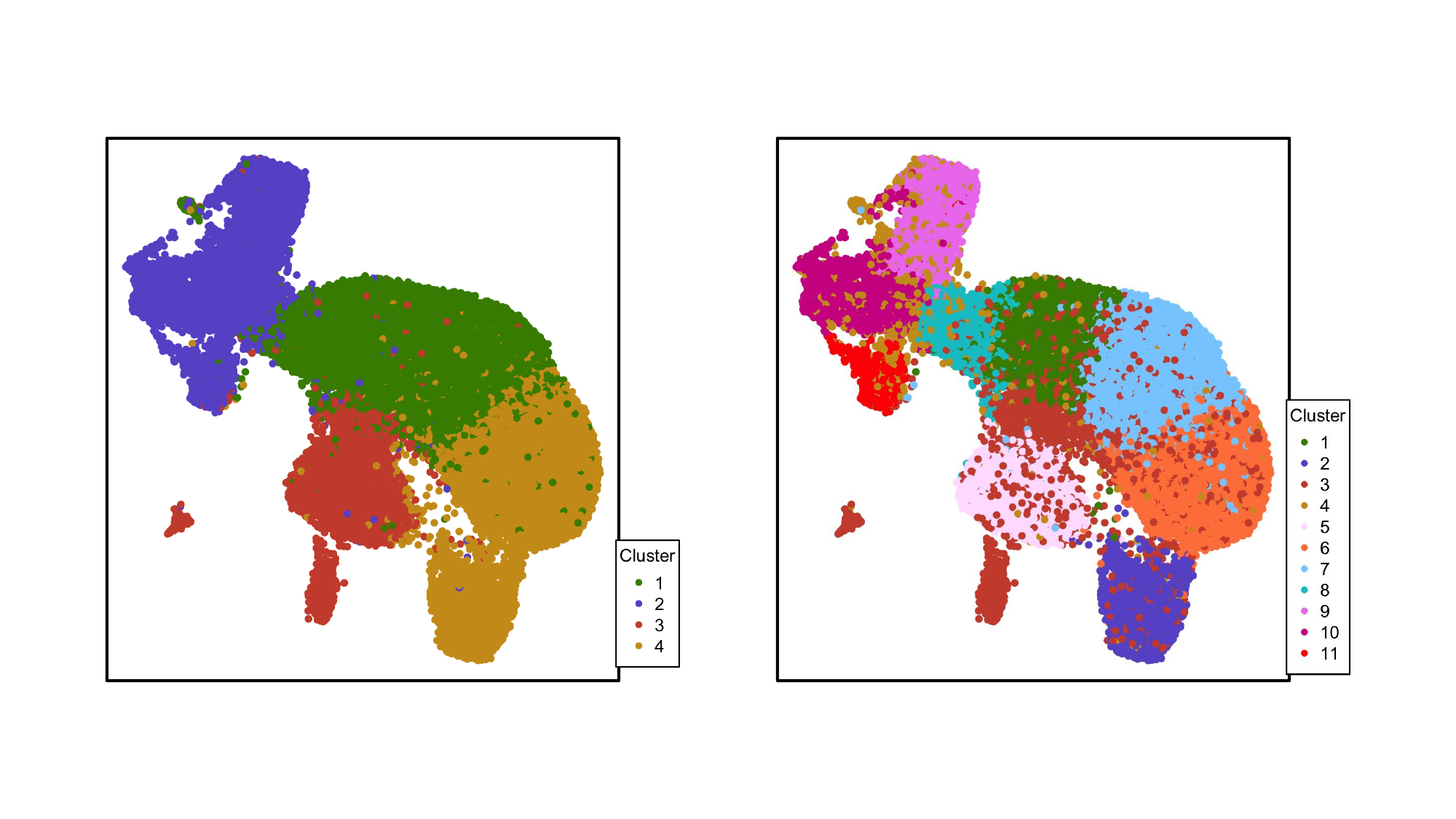}
    \caption{\textbf{The visualisation plots of different $K$ for GMM in UMAP-LE projection.} $K=4$ (left); $K=11$ (right)}
    \label{fig:GMM gw1722 c1}
\end{figure}

\begin{figure}[H]
    \centering
    \includegraphics[width = 0.97\textwidth]{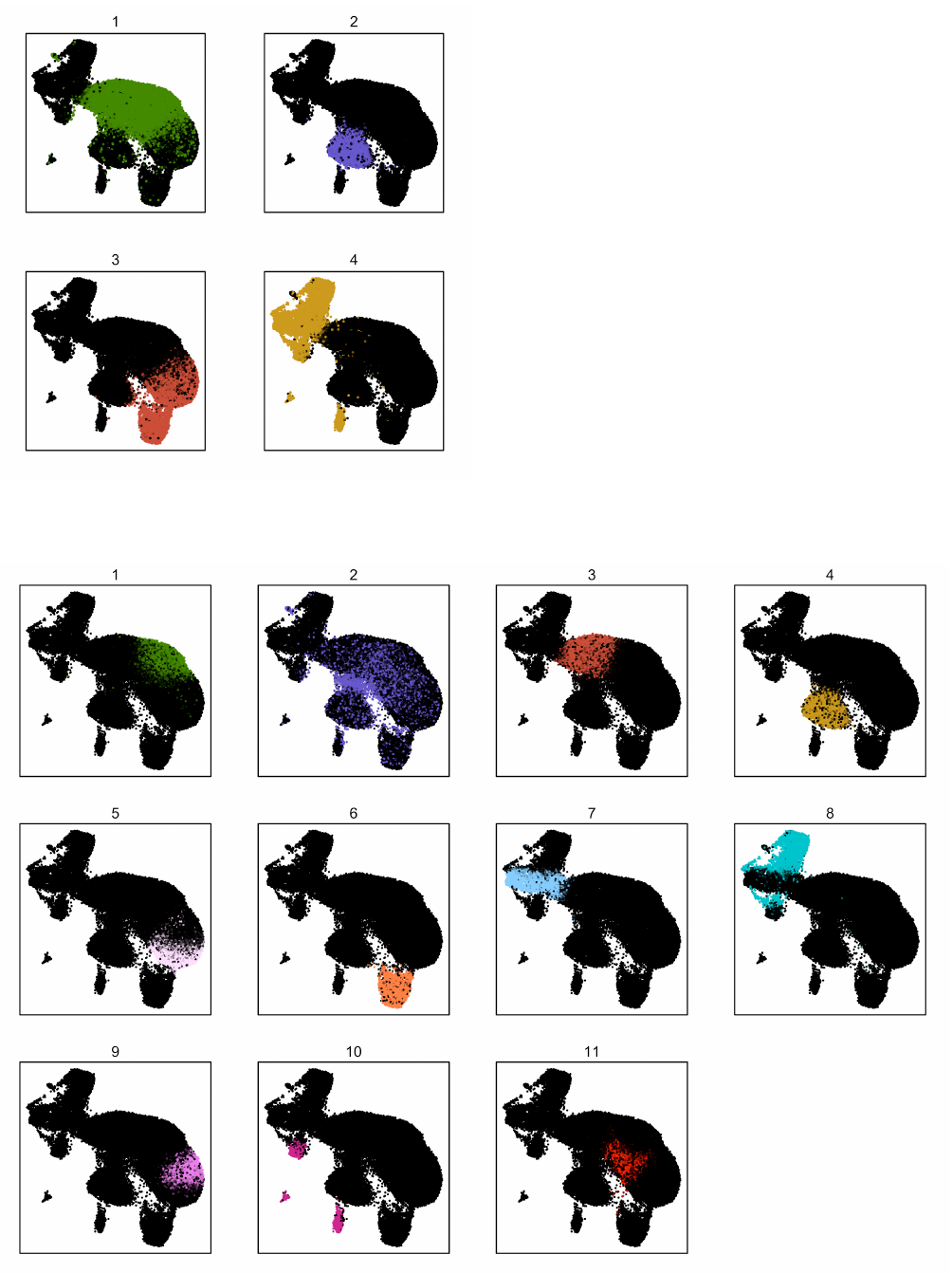}
    \caption{\textbf{The visualisation plots of different $K$ for VB-GMM in UMAP-LE projection.} $K=4$ (upper); $K=11$ (bottom)}
    \label{fig:VI gw1722 c1}
\end{figure}


\begin{table}[htbp]
\centering
\begin{tabular}{l|r|r|r|r}
 Gene & logFC & AveExpr &    $t$ &      $B$\\
\hline
       KLF6* & 2.090 &   7.787 & 60.871 & 1421.039 \\
       EMP1* & 1.776 &   7.153 & 58.890 & 1348.153 \\
       RTN4* & 1.551 &   7.991 & 57.559 & 1299.243 \\
      DDX21* & 1.355 &   8.280 & 52.059 & 1101.404 \\
      YWHAH* & 1.584 &   7.837 & 51.872 & 1094.870 \\
      CYTOR* & 1.454 &   7.054 & 51.502 & 1081.872 \\
       EIF1* & 1.195 &   9.823 & 51.034 & 1064.914 \\
       ATF4* & 1.337 &   8.077 & 50.575 & 1049.242 \\
     TM4SF1* & 1.858 &   8.534 & 50.341 & 1040.964 \\
    LRRFIP1* & 1.364 &   7.373 & 50.238 & 1037.543 \\
      YWHAZ* & 1.374 &   8.020 & 49.874 & 1024.765 \\
       TPM4* & 1.457 &   8.163 & 49.135 &  999.121 \\
     TXNRD1* & 1.307 &   6.975 & 49.098 &  997.929 \\
        FTL* & 1.660 &   9.629 & 48.871 &  989.718 \\
      CYR61* & 1.662 &   7.155 & 48.772 &  986.678 \\
     SLC3A2* & 1.311 &   7.146 & 48.377 &  973.074 \\
MIR4435-2HG* & 1.259 &   6.650 & 48.356 &  972.370 \\
        EZR* & 1.353 &   6.950 & 47.950 &  958.423 \\
      ACTN4* & 1.316 &   7.119 & 47.796 &  953.142 \\
      ITGB1* & 1.323 &   7.592 & 47.398 &  939.550 \\
      FOSL1* & 1.179 &   6.706 & 46.906 &  922.852 \\
      ITGA2* & 1.330 &   6.698 & 46.779 &  918.541 \\
      CLIC1* & 1.281 &   8.426 & 46.396 &  905.500 \\
     PDLIM5* & 1.290 &   7.180 & 46.180 &  898.302 \\
    SELENOS* & 1.168 &   7.247 & 46.025 &  893.095 \\
\end{tabular}
\caption{\textbf{DEA results for the breast cancer data using VB-GMM} For $K=8$, cluster $4$ (control) and cluster $7$ (treatment) were identified for analysis. \textit{*} indicates genes with both $P$-value $< 0.001$ and adjusted $P$-value $< 0.001$.}
\label{tab:Top25DEA_br}
\end{table}

\begin{table}[htbp]
\centering
\begin{tabular}[]{l|r|r|r|r}
 Gene & logFC & AveExpr &    $t$ &      $B$\\
\hline
       TNFAIP6* & 3.199 & 9.008 & 83.374 & 2376.685 \\
       TM4SF1* & -4.524 & 7.566 & -82.596 & 2344.149 \\
       HLA-B* & -2.992 & 8.000 & -77.351 & 2129.65 \\
      B2M* & -2.357 & 10.240 & -77.303 & 2126.928 \\
      EIF1* & -1.979 & 9.996 & -73.893 & 1988.042 \\
      ADAMTS9* & -2.934 & 6.628 & -72.346 & 1925.886 \\
       TPT1* & -2.088 & 10.034 & -70.986 & 1869.933 \\
       PNP* & -2.267 & 6.505 & -70.089 & 1834.385 \\
     EEF1A1* & -1.820 & 11.026 & -67.150 & 1714.955 \\
     DCN* & 2.066 & 7.587 & 66.824 & 1702.793 \\
      MALAT1* & -1.925 & 12.739 & -66.588 & 1692.129 \\
       CLMP* & 1.395 & 6.920 & 65.468 & 1648.389 \\
     RHOB* & -2.452 & 6.694 & -64.941 & 1627.439 \\
     TCF4* & -2.330 & 6.510 & -64.184 & 1597.299 \\
      PTMA* & -1.725 & 10.205 & -64.061 & 1591.879 \\
     HLA-C* & -2.463 & 7.486 & -63.860 & 1584.367 \\
    CD74* & -2.440 & 6.422 & -63.500 & 1570.129 \\
     H3F3B* & -1.608 & 9.720 & -62.783 & 1541.392 \\
      H2AFZ* & -2.380 & 7.562 & -62.472 & 1529.374 \\
      NEAT1* & -2.664 & 9.131 & -62.293 & 1522.05 \\
      FABP4* & -2.934 & 6.640 & -62.199 & 1518.653 \\
      GADD45B* & -2.446 & 6.958 & -62.195 & 1518.471 \\
      LUM* & 1.414 & 6.803 & 61.079 & 1474.358 \\
     TMSB10* & -2.644 & 9.160 & -60.718 & 1460.111 \\
    ACTG1* & -2.831 & 7.724 & -60.416 & 1448.524 \\
\end{tabular}
\caption{\textbf{DEA results for the breast cancer data using GMM} For $K=8$, cluster $7$ (control) and cluster $3$ (treatment) were identified for analysis. \textit{*} indicates genes with both $P$-value $< 0.001$ and adjusted $P$-value $< 0.001$.}
\label{tab:Top25DEA_br_GMM}
\end{table}

\begin{table}[htbp]
\makebox[\textwidth][c]{%
\begin{tabular}{l|r|r|r|r}
   & {logFC} & {AveExpr} & {t} & {B} \\
  \hline
     FABP7* & -2.461 & 7.767 & -56.145 & 1104.588 \\
    DOK5* & -1.328 & 6.548 & -36.981 & 567.927 \\
    LMO4* & -1.350 & 6.710 & -33.271 & 471.636 \\
    SATB2* & -0.893 & 6.244 & -30.636 & 407.356 \\
    BHLHE22* & -0.767 & 6.190 & -28.525 & 357.574 \\
    NPY* & -0.800 & 6.257 & -26.018 & 301.217\\
\end{tabular}
}
\end{table}
\begin{table}[htbp]
\resizebox{\textwidth}{!}{
\begin{tabular}{l|r|r|r}
  &  {OR} & {95\% CI Lower} & {95\% CI Upper} \\
\hline
    ZHONG\_PFC\_C3\_MICROGLIA* & 15.395 & 9.789 & 23.925 \\
    FAN\_EMBRYONIC\_CTX\_BIG\_GROUPS\_EXCITATORY\_NEURON* & 66.042 & 29.407 & 145.951 \\
    DESCARTES\_FETAL\_LUNG\_VISCERAL\_NEURONS & 18.092 & 9.947 & 31.614 \\
    ZHONG\_PFC\_C8\_UNKNOWN\_NEUROD2\_POS\_INTERNEURON* & 27.615 & 13.442 & 53.794 \\
    MANNO\_MIDBRAIN\_NEUROTYPES\_HGABA* & 6.539 & 4.133 & 10.173 \\
\end{tabular}
}
\caption{\textbf{DEA and GSEA results for clusters $5$ and $6$ in neuron data} Results were obtained using VB-GMM with $K=14$, where cluster $5$ is control and cluster $6$ is treatment. The top table shows negative outcomes from DEA; The bottom table shows negative outcomes from GSEA. \textit{*} indicates genes with both $P$-value $< 0.001$ and adjusted $P$-value $< 0.001$}.
\label{tab:GSEA_DEA_SIG}
\end{table}

\begin{table}[htbp]
\makebox[\textwidth][c]{%
\begin{tabular}{l|r|r|r|r}
  &  logFC & AveExpr &       t &    B \\
  \hline
     DLX6-AS1* & -0.999 & 6.671 & -49.174 & 1065.231 \\
      PLS3* & -0.511 & 6.183 & -37.649 & 651.053 \\
    RBP1* & -0.493 & 6.278 & -35.193 & 572.788 \\
   CD109* & -0.005 & 5.787 & -34.571 & 552.071\\
     TEK* & -0.005 & 5.787 & -34.346 & 545.172\\
     DLX5* & -0.491 & 6.165 & -34.255 & 544.027\\
\end{tabular}
}
\end{table}
\begin{table}[htbp]
\resizebox{\textwidth}{!}{
\begin{tabular}{l|r|r|r}
  &  {OR} & {95\% CI Lower} & {95\% CI Upper} \\
\hline
    ZHONG\_PFC\_MAJOR\_TYPES\_INTERNEURON* & 298.547 & 88.698 & 1293.373\\
    ZHONG\_PFC\_C6\_DLX5\_GAD1\_GAD2\_POS\_INTERNEURON* & 245.031 & 68.832 & 1051.036\\
    FAN\_EMBRYONIC\_CTX\_BIG\_GROUPS\_INHIBITORY* & 383.003 & 74.807 & 4037.517\\ ZHONG\_PFC\_C7\_SST\_LHX6\_POS\_PUTATIVE\_MIGRATING\_INTERNEURON* & 127.232 & 37.941 & 459.835\\
    FAN\_EMBRYONIC\_CTX\_IN\_4\_INTERNEURON* & 219.891 & 49.461 & 1353.407\\
  
\end{tabular}
}
\caption{\textbf{DEA and GSEA results for clusters $4$ and $12$ neuron data.} Results were obtained using VB-GMM with $K=14$, where cluster $4$ is control and cluster $12$ is (
treatment. The top table shows negative outcomes from DEA; The bottom table shows negative outcomes from GSEA. \textit{*} indicates genes with both $P$-value $< 0.001$ and adjusted $P$-value $< 0.001$.}
\label{tab:GSEA_DEA_SIG1}
\end{table}

\end{appendices}

\end{document}